\documentclass[twoside]{article}
\usepackage{graphicx}
\usepackage{fancyhdr}
\usepackage{multicol}
\usepackage{styl-ap}

\begin{document}
\title{Data Analysis of Globular Cluster Harris Catalogue in view of the King models and their dynamical evolution. II. Observational Evidences.}
\author{Marco Merafina\work{1}, Daniele Vitantoni\work{1}}
\workplace{Department of Physics, University of Rome La Sapienza, Piazzale Aldo Moro 2, I-00185 Rome, Italy}
\mainauthor{marco.merafina@roma1.infn.it}
\maketitle

\begin{abstract}
We summarize some observational comparison concerning the features of 
globular clusters (GCs) population in connection to the evolution of King 
models. We also make a comparison with some extragalactic GCs systems, in 
order to underline the effects of the main body on the dynamical evolution.
\end{abstract}


\begin{multicols}{2}

\section{Introduction}

Globular clusters (GCs), for their proprieties of symmetry and their high 
relaxation times, are important to test theories about thermodynamical 
stability of spherical self gravitating systems. The actual sample of 
is a mixture of various and not homogeneus GCs types. Therefore, it is difficult to analyze properties of Milky Way (MW) GCs population in connection to core-collapse and gravotermal instability.

The last version (2010) of Harris GCs Catalogue (see also Harris, 1996) includes 157 objects. It was pointed out by van Der Bergh (2011) that Harris catalogue could includes three not typical GCS, probably remnant cores of DSph galaxies: Omega Centauri, Terzan 5 and NGC 6715 (M54). The Harris catalogue also includes PCC GCs, namely GCs with collapsed cores that cannot be described by classical single mass King models profiles.

Zinn (1985) identifies two classes of GCs, respectively known as disk 
population (metal rich) and halo population (metal poor), distinct by the
threshold value $[Fe/H]\simeq -0.8$ (or, according to some authors, 
$-0.75$).

Recently Bica et al. (2006) showed that the actual GCs population 
seems to have been contaminated by capture of smaller galaxies (and their 
possible GCs populations) during the Milky Way formation. Possible 
evidences of extragalactic origin of some GCs are retrograde motion 
(compared to galactic disk motion) and unusual young absolute age.

It seems that the original GCs population suffered deep and incisive 
processes of disruption (see Aguilar et al., 1988; Hut \& Djorgovski, 1992; Gnedin \& Ostriker, 1997; Mackey \& Gilmore, 2004), until almost 50\% of original GCs are destroyed in the last Hubble time.

\section{Discussion}

We start to consider the problem introduced by Katz in the paper about 
thermodynamic stability in 1980. The study of the distribution of galactic 
GCs in terms of $W_{0}$ (central gravitational potential) shows a peak 
value of 6.9. We should expect that the peak value coincides with the 
known stability critical value $W_{0}=7.4$, due to the old age of MW GCs 
and the onset of the instability in the high $W_{0}$ region. This problem 
had remained unsolved (Fig.\ref{fig1}).

\begin{myfigure}
\centerline{\resizebox{70mm}{!}{\includegraphics{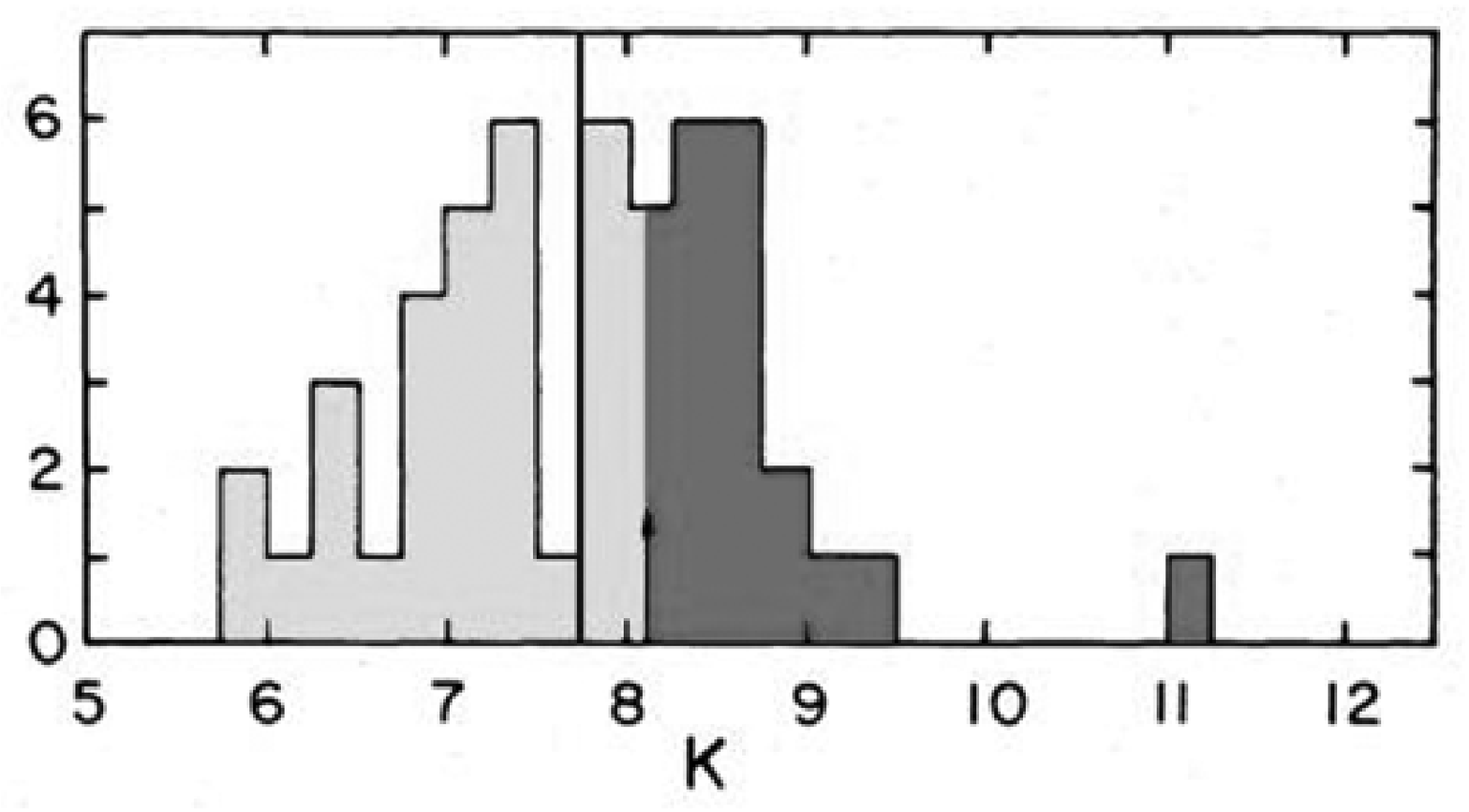}}}
\caption{Distribution of galactic GCs at different $K$ (Katz, 1980). The quantity $K$ is related to $W_0$ (see Merafina \& Vitantoni, Part I).}
\label{fig1}
\end{myfigure}

With the introduction of the effective potential (see Merafina \& Vitantoni, Part I) and including the additional term in the expression of the total energy, we can revise the Katz study. The result is a very satisfactory coincidence of the observative peak value with the stability limit. We can also repeat the analysis on a more detailed and updated sample (using data of the Harris GCs Catalogue). The peak value, in the non-symmetric Gaussian hypothesis, is exactly at $W_{0}=6.9$ (Fig.\ref{fig2}).

\begin{myfigure}
\centerline{\resizebox{85mm}{!}{\includegraphics{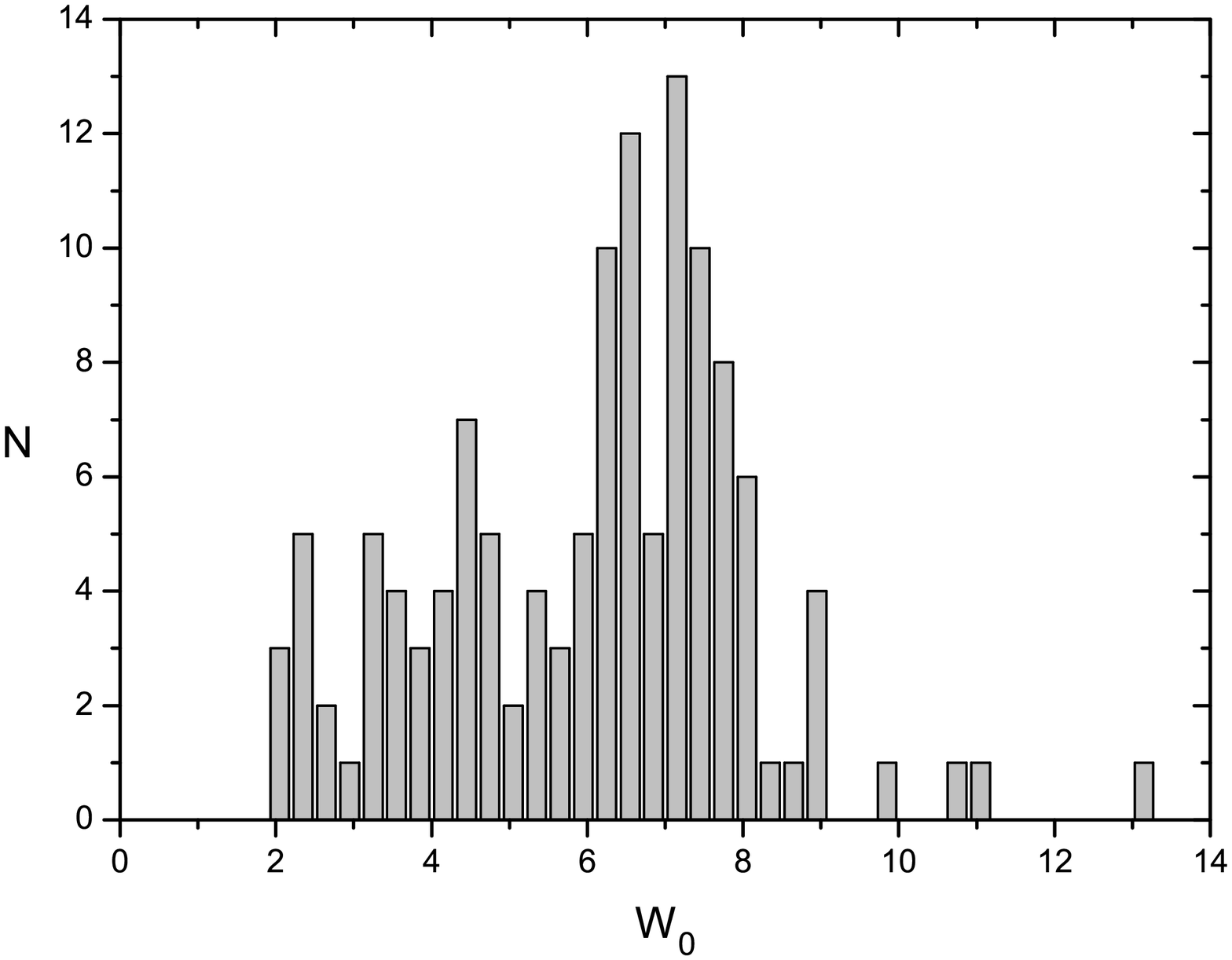}}}
\caption{$W_{0}$ distribution of pre-core-collapse MW GCs.}
\label{fig2}
\end{myfigure}
For a better understanding of the evolution of a GCs population, we 
briefly analyze the role of environmental features. The effect of the 
distance from the Galactic center (Fig.\ref{fig3}) is not clear at all. 
Generally speaking, the more a GC lives near the galactic center, the more 
quickly it evolves towards the gravothermal catastrophe, being more affected by tidal forces of the Galaxy.
\begin{myfigure}
\centerline{\resizebox{85mm}{!}{\includegraphics{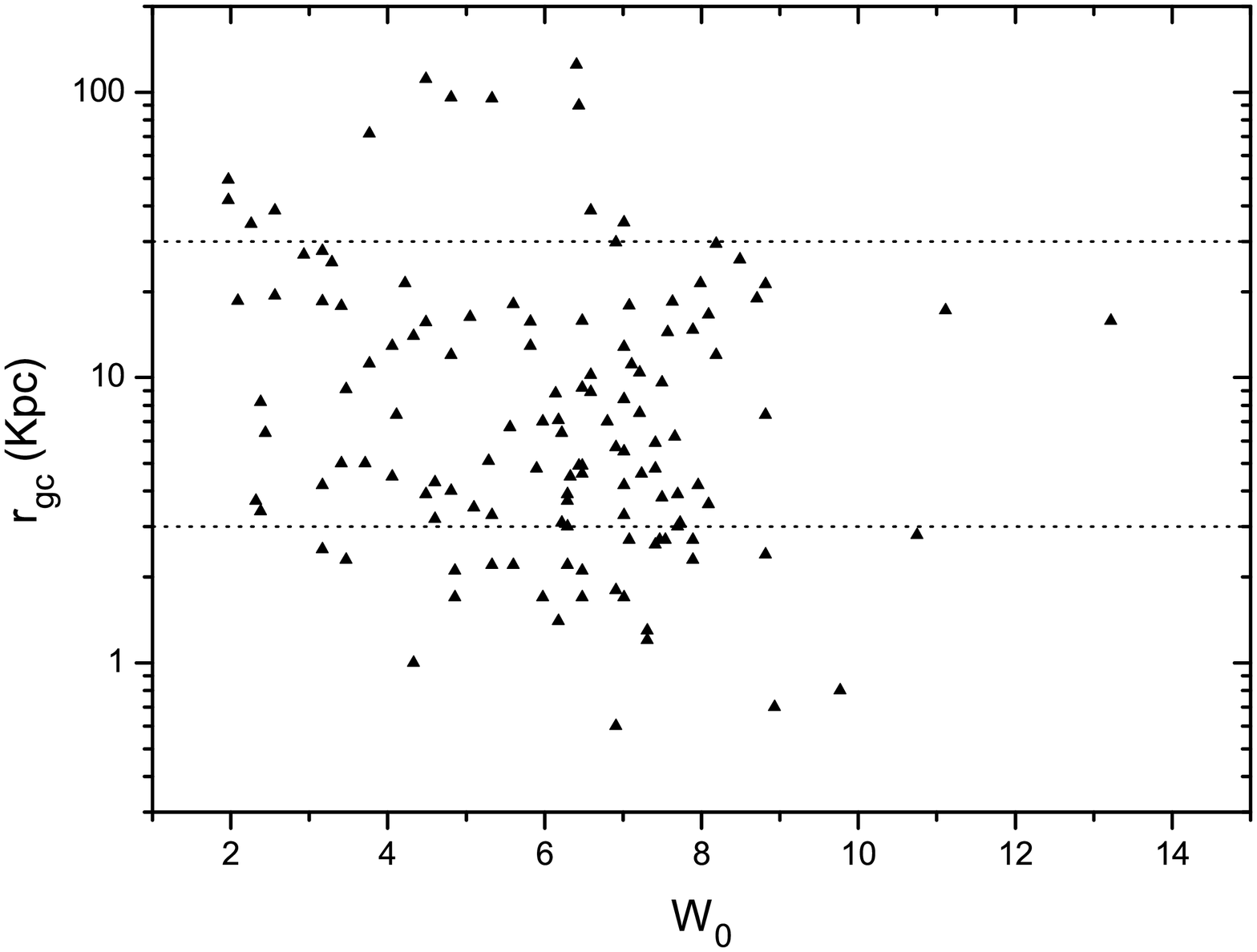}}}
\caption{Galactocentric distance $r_{gc}$ in function of $W_{0}$. The dashed lines represent the values at 3Kpc and 30Kpc.}
\label{fig3}
\end{myfigure}
If we look at the GCs distribution in the $[Fe/H]$-$W_{0}$ plane 
(Fig.\ref{fig4}), we find no correlations between these two quantities. 
This means that the difference between halo and disk population, first 
introduced by Zinn (1985) does not influence the dynamical features and 
the evolution. Nevertheless, if we analyze the disk population 
(Fig.\ref{fig5}), this seems to be more dynamically evolved than the halo 
one. The $W_0$ peak value is slightly larger for the disk population, mainly due to a lower $r_{gc}$ (in average) for this class of objects. On the other hand, it is well known that the tidal shocks played a more incisive role in the evolution of the disk GCs.

\begin{myfigure}
\centerline{\resizebox{90mm}{!}{\includegraphics{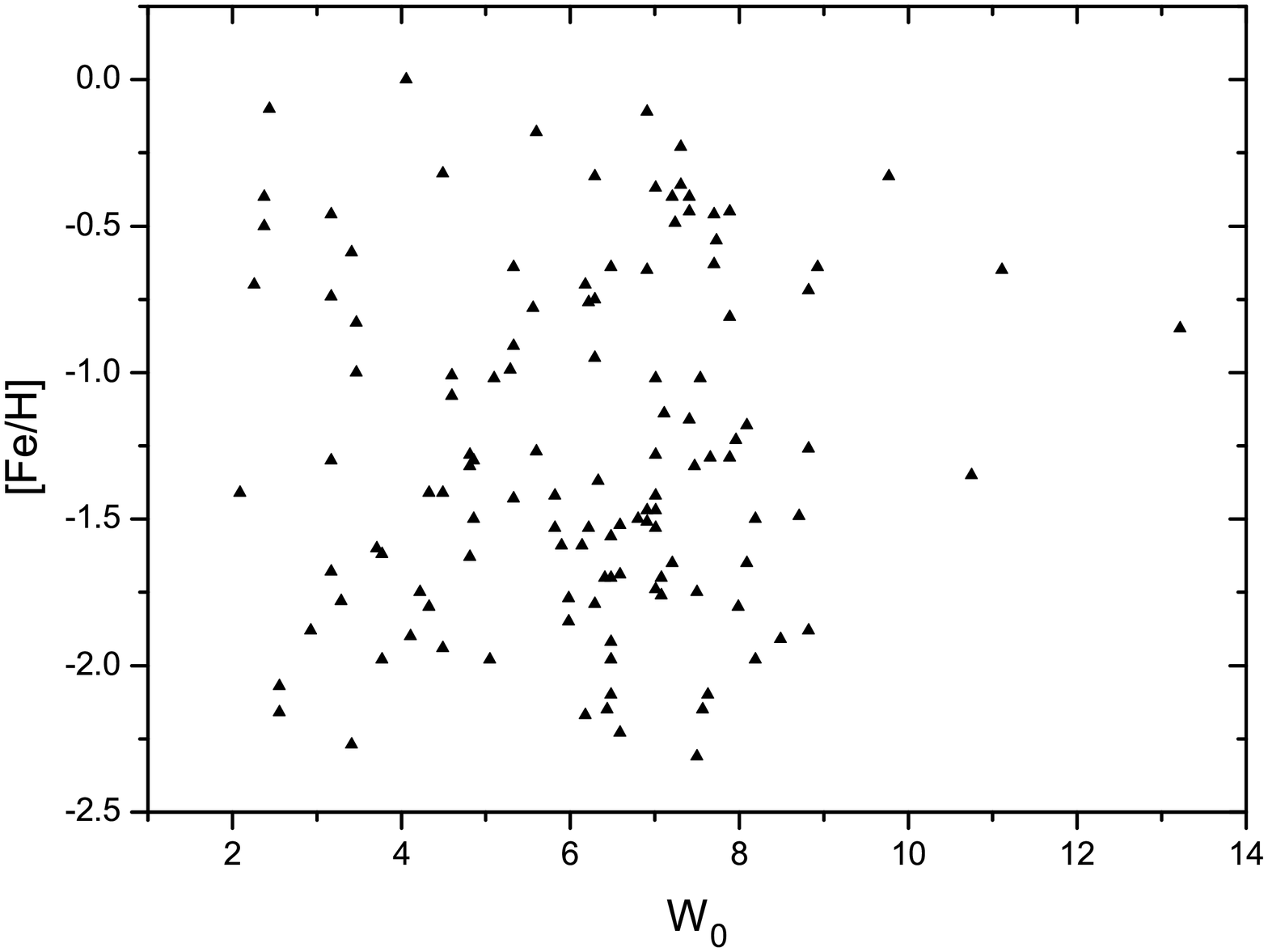}}}
\caption{Total metallicity $[Fe/H]$ in function of $W_{0}$.}
\label{fig4}
\end{myfigure}

\begin{myfigure}
\centerline{\resizebox{90mm}{!}{\includegraphics{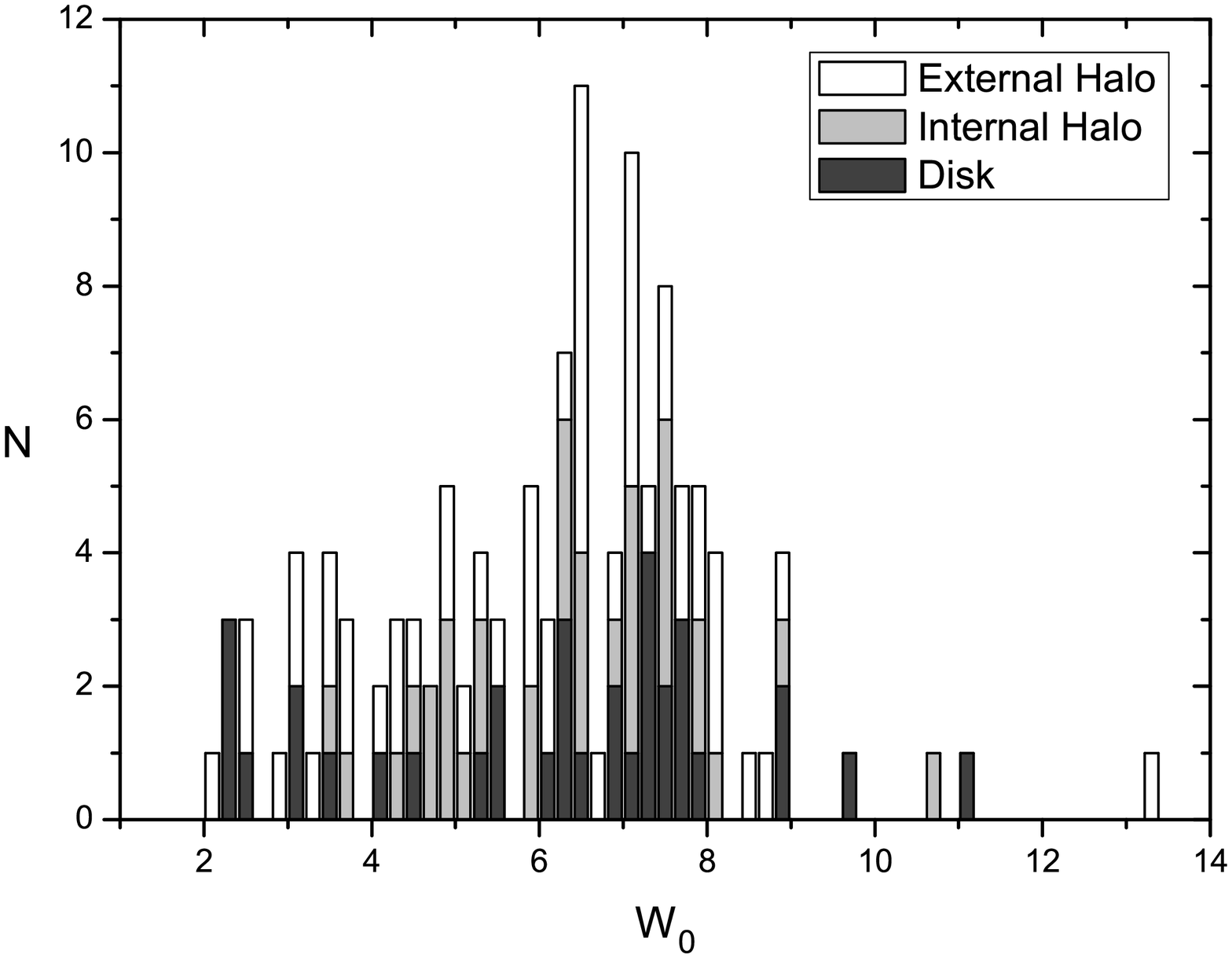}}}
\caption{$W_{0}$ distribution of astronomical populations.}
\label{fig5}
\end{myfigure}

\begin{myfigure}
\centerline{\resizebox{70mm}{!}{\includegraphics{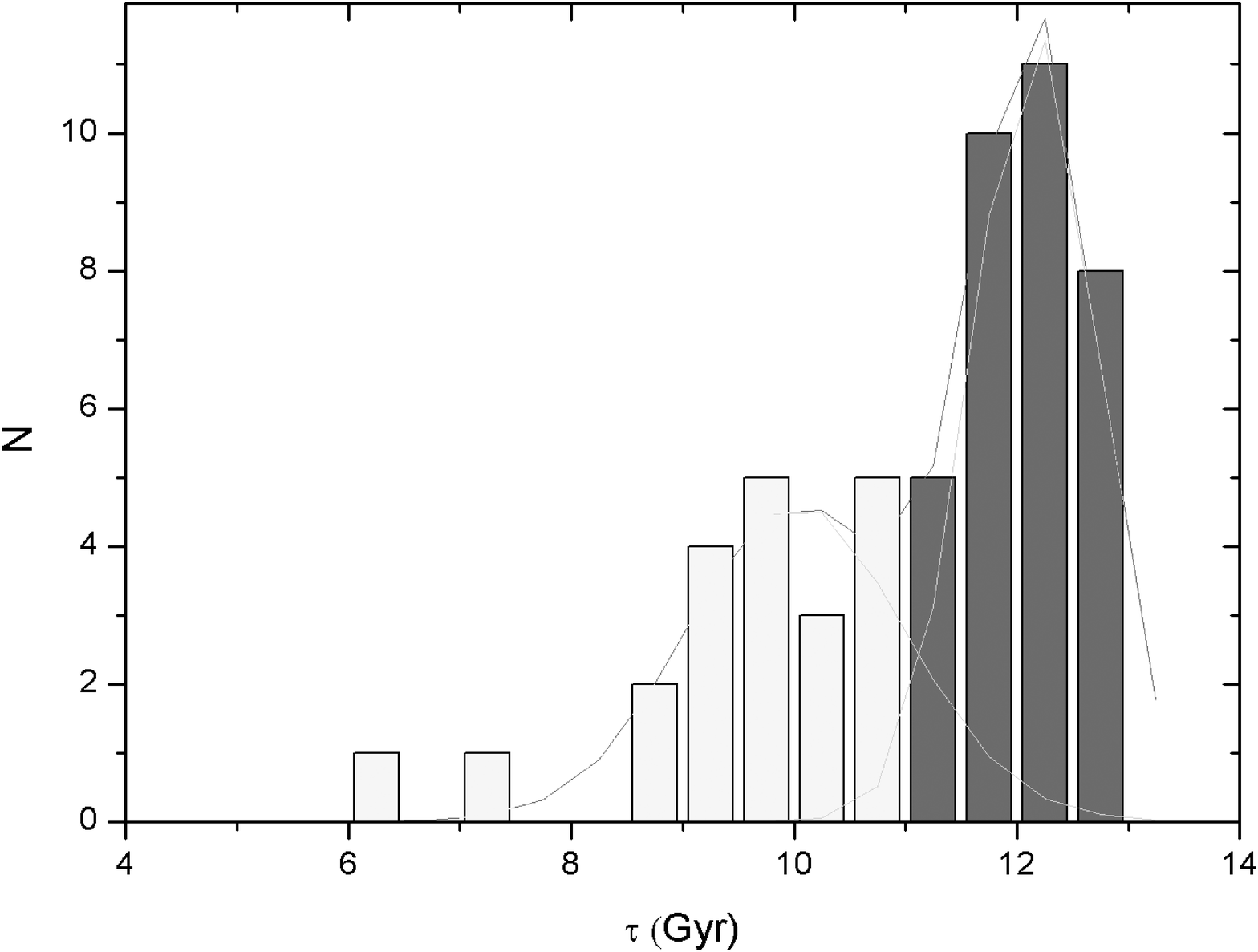}}}
\caption{Absolute age distribution for MW GCs. Older GCs, presumably all native in the Milky Way, are evidenced in dark grey.}
\label{fig6}
\end{myfigure}

The contamination of the Halo GCs with the extragalactic origin ones is 
suggested by the bimodality in the absolute age distribution, by the 
rotation in the plane of Galactic Disk, and by the Age-Metallicity 
dependence. The situation is shown in Figs.\ref{fig6}, \ref{fig7} and \ref{fig8}. Regarding the Age-Metallicity dependence, we can say that few objects are located out of the main well defined sequence of clusters (and probably have a different origin with respect to all the others).

\begin{myfigure}
\centerline{\resizebox{70mm}{!}{\includegraphics{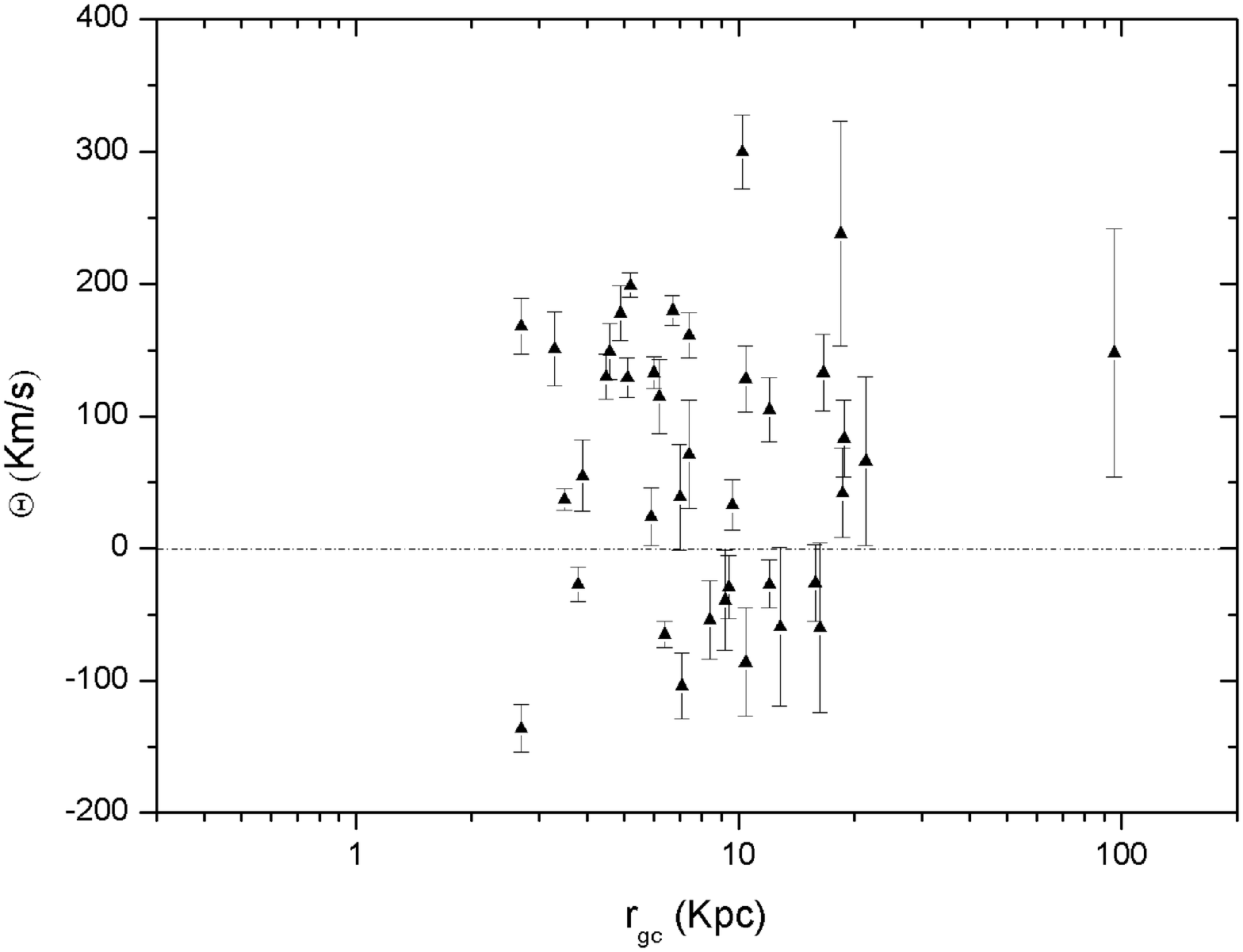}}}
\caption{Motion over the Galactic plane for MW GCs (from Dinescu et al., 1999).}
\label{fig7}
\end{myfigure}

\begin{myfigure}
\centerline{\resizebox{90mm}{!}{\includegraphics{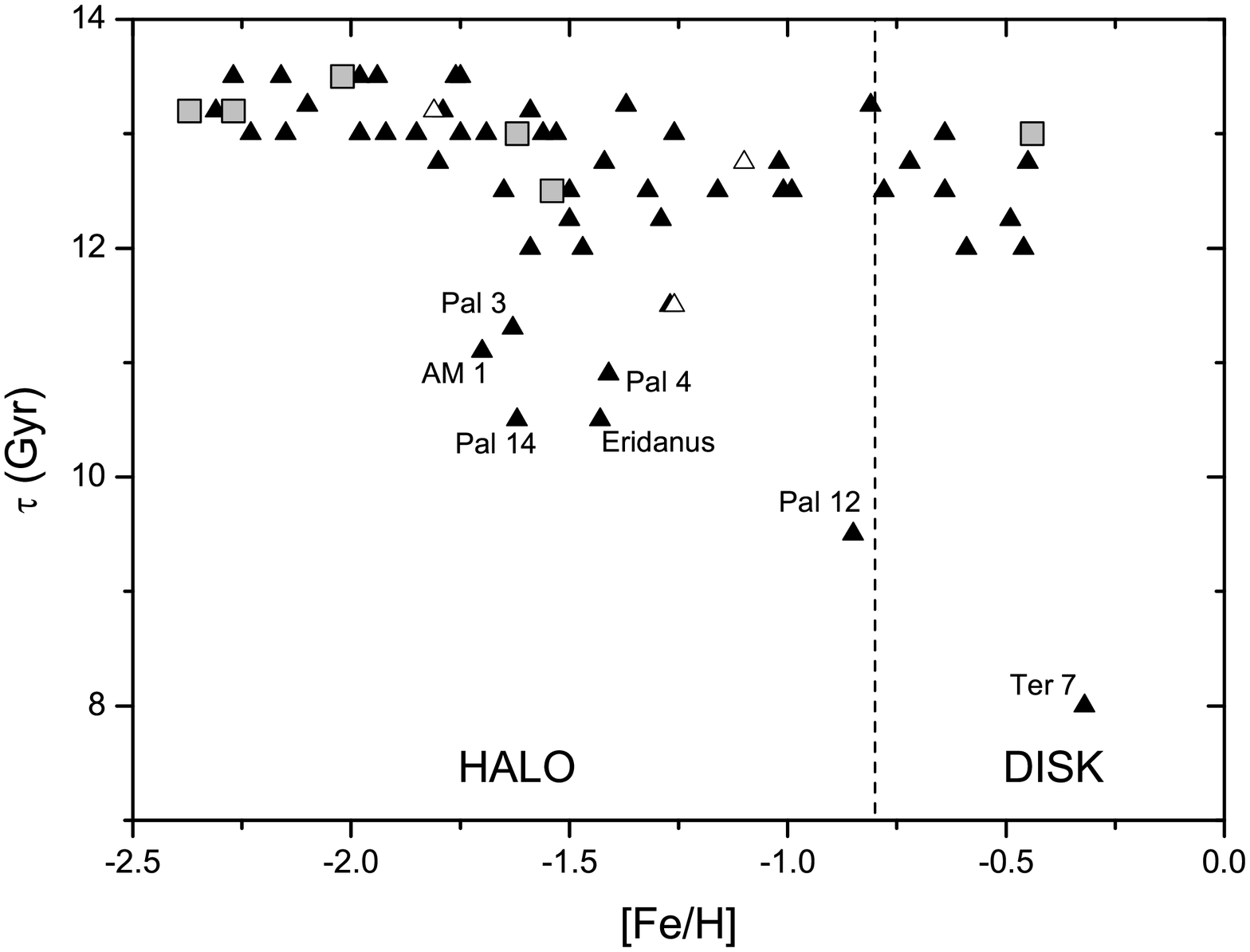}}}
\caption{Age-Metallicity behavior, using the absolute age values extimated 
by Dotter et al. (2010). PCC clusters are indicated by grey squares, suspected PCC ones by white triangles.}
\label{fig8}
\end{myfigure}

If we analyze the behavior of the central relaxation time $t_{rc}$ in 
function of $W_{0}$ (Fig.\ref{fig9}), we can note a linear decreasing, 
that indicates an increasing of evolutional speed towards the collapse. 
Below the treshold value $\log t_{rc} = 8.0$ there is a region of 
coexistence of Pre-Core Collapse GCs with Post-Core Collapse (PCC) ones. 
The behavior of core collapse time $t_{cc}$ (that is the remaining time 
before the collapse of the model), changes over the stability limit 
$W_{0}=6.9$ (Fig.\ref{fig10}).
\begin{myfigure}
\centerline{\resizebox{90mm}{!}{\includegraphics{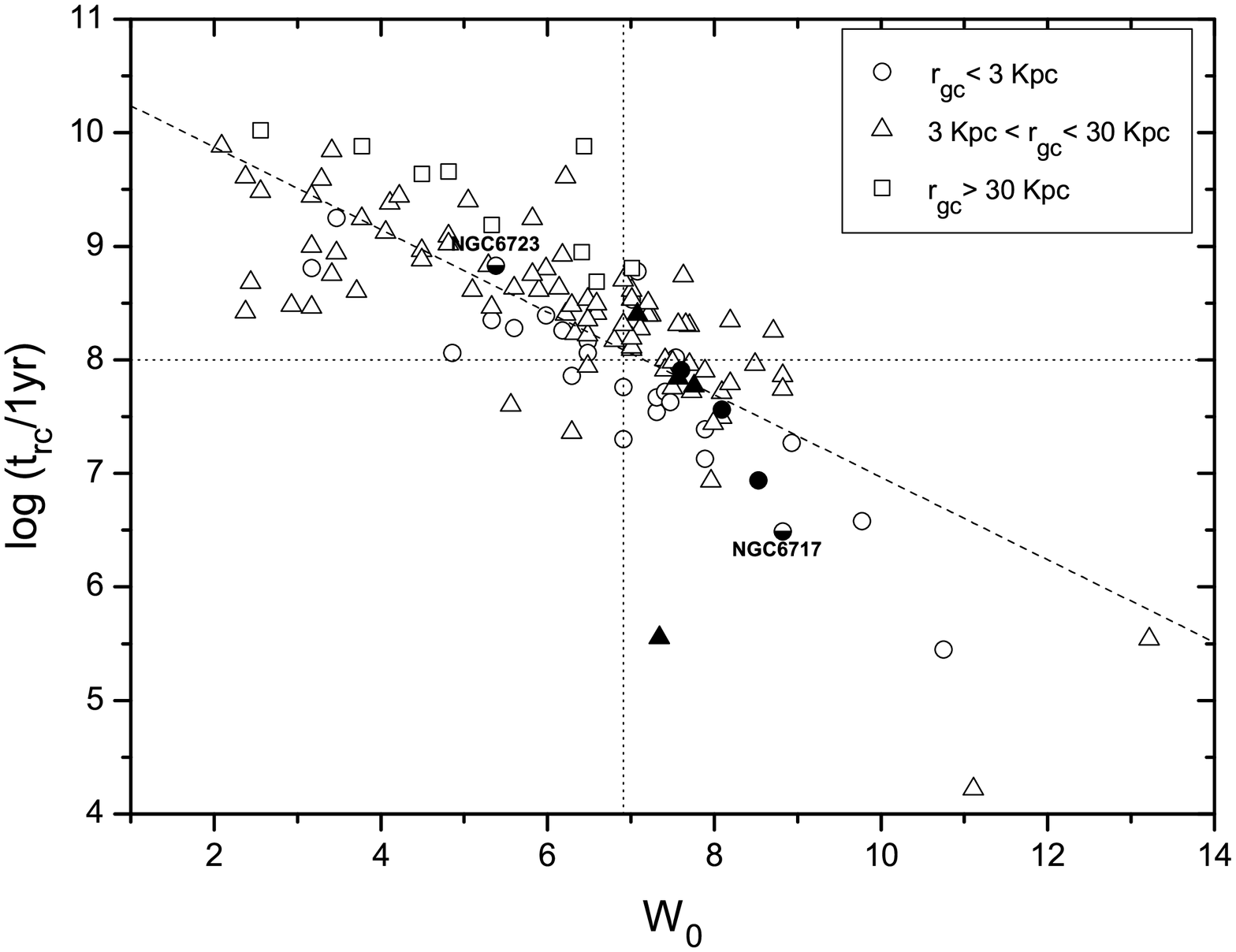}}}
\caption{Behavior of $t_{rc}$ in function of $W_{0}$; three classes of 
cluster distances are represented. Vertical dashed line represent the 
stability limit, whereas the horizontal one represent the $t_{rc}$ critical 
value (Cohn \& Hut, 1984) distinguishing \emph{pre-core-collapsed} and 
\emph{post-core-collapsed} objects. Suspected PCC clusters are indicated by filled symbols. In the Harris Catalogue NGC6723 has been erroneously included among the suspected PCC clusters in place of NGC6717.}
\label{fig9}
\end{myfigure}
\begin{myfigure}
\centerline{\resizebox{90mm}{!}{\includegraphics{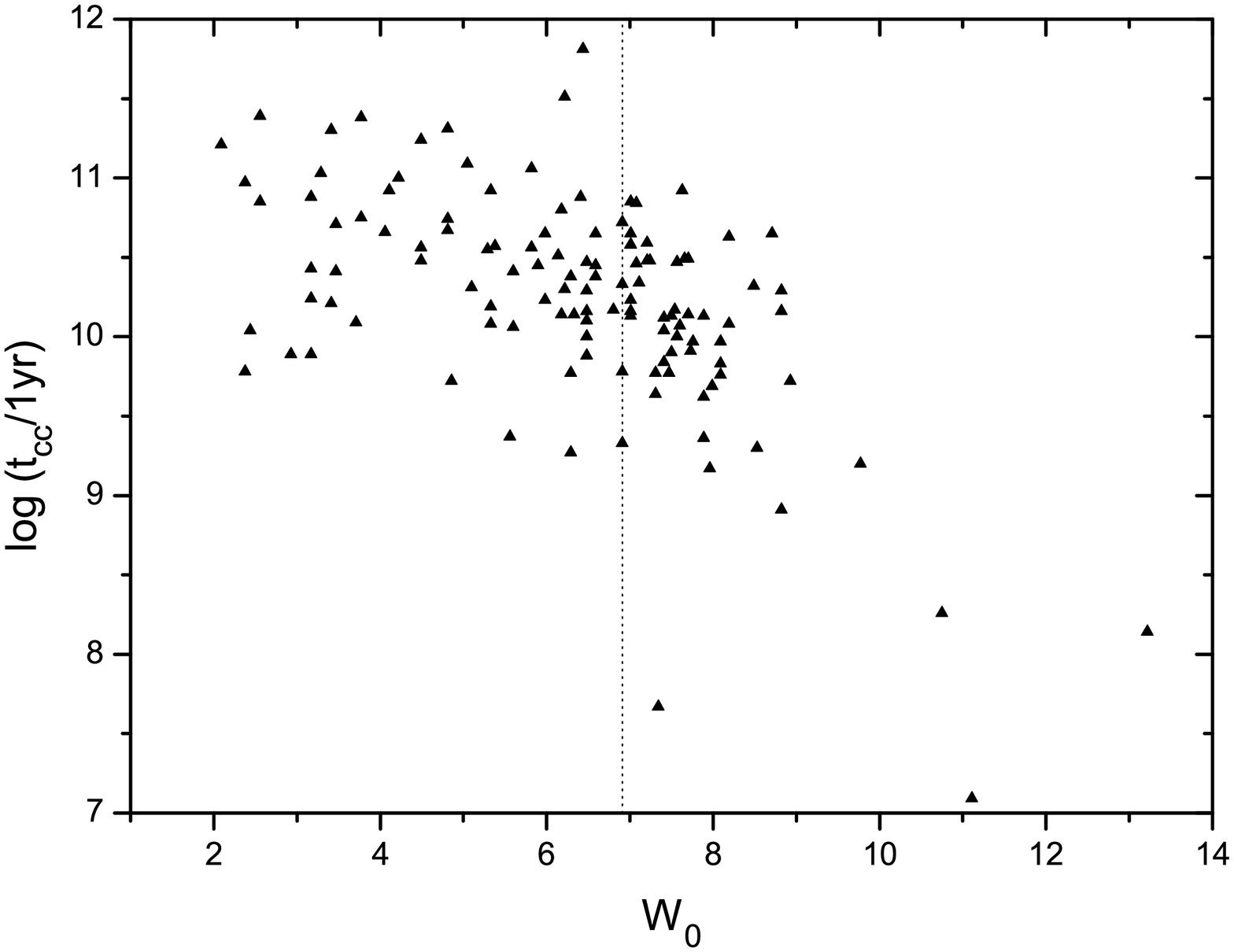}}}
\caption{Core-collapse time in function to $W_{0}$. The $t_{cc}$ values 
are estimated as in Quinlan (1996).}
\label{fig10}
\end{myfigure}

We also consider a comparison among four GCs populations, shown in 
Fig.\ref{fig11}: the LMC is the less evolved, as well as the SMC and Fornax systems (Mackey \& Gilmore, 2003b, 2003c); it presents a peak value 
close to $W_{0}=4.3$. For this kind of GCs population, the presence of a low massive main body allowed to preserve more informations about primeval distribution features. On the contrary, NGC5128, whose main body is a giant elliptical galaxy, seems to be an evolved population with a maximum 
value up to the threshold value $W_{0}=6.9$. Finally, the M31 system is the most similar to our GCs population, with a main peak value around the stability limit and an extended tail in low-$W_{0}$ region, that is produced by the presence of low evolutionary speed objects or extragalactic origin clusters. Also the effects of disk and bulge shocking, realistically, concurred to the formation of the tail.
\begin{myfigure}
\centerline{\resizebox{80mm}{!}{\includegraphics{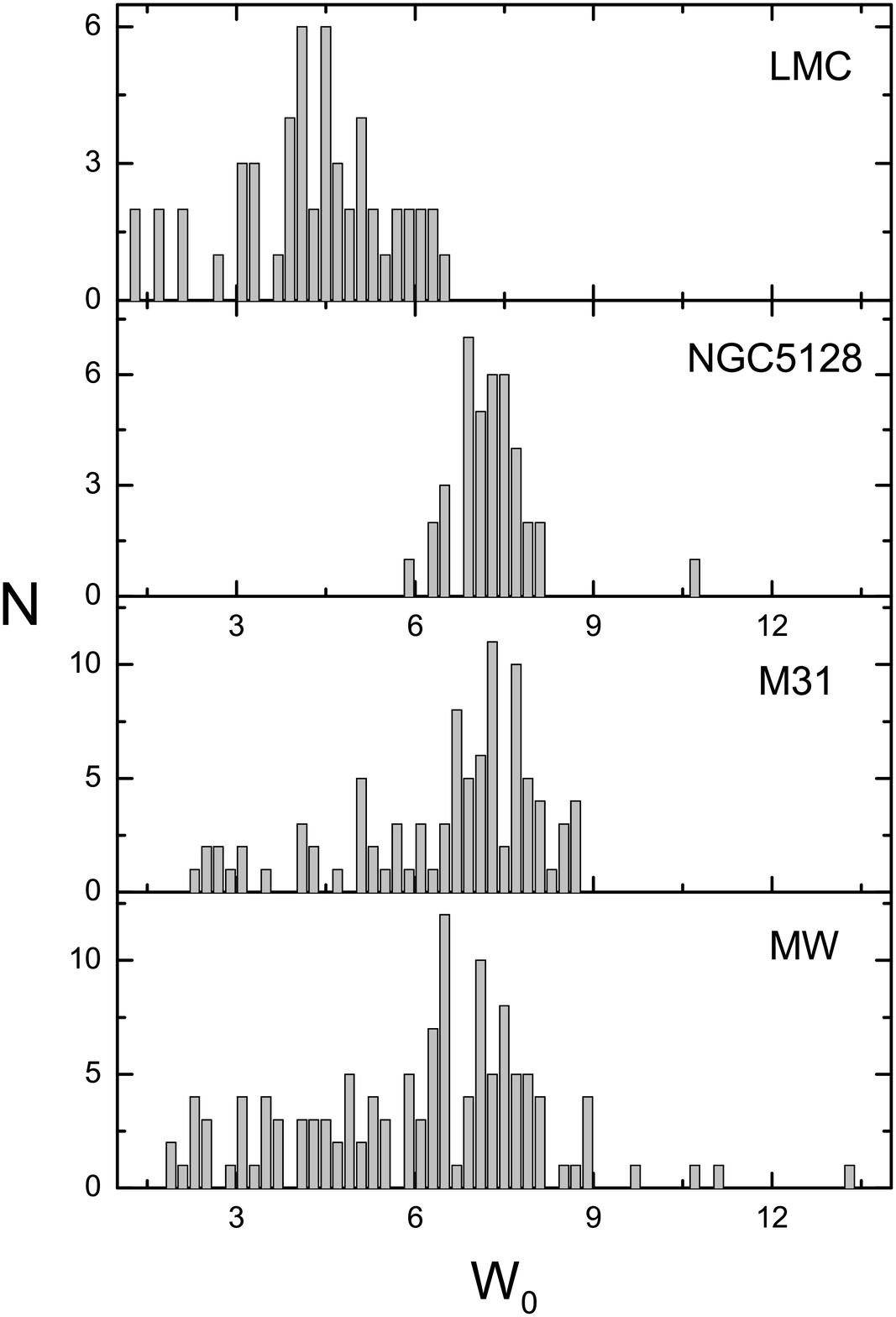}}}
\caption{Comparison among $W_{0}$ distributions of different GCs 
systems. In addition to MW system, M31, NGC5128 and LMC systems are reported. M31 data come (or are deduced) from Barmby et al. (2007), NGC5128 ones from G\`omez et al. (2005), LMC data from Mackey \& Gilmore (2003a). MW histogram differences from Fig.2 are given by a different binning choice.}
\label{fig11}
\end{myfigure}

\section{Conclusions}
In order to analyze the dynamical evolution of King single mass GCs, we 
have analyzed the MW GCs population. The MW clusters $W_{0}$ 
distribution presents a peak very close to the new stability limit 
$W_{0}=6.9$ and a pronounced tail in the low-$W_{0}$ region.

We can instead exclude a direct relation between astronomical GCs 
populations and dynamical evolution, except for a very weak increasing of 
$W_{0}$ peak value for disk clusters.

From the time-scales we can deduce that clusters with high $W_{0}$ value 
have an higher collapsing speed. The gravothermal catastrophe produce an 
alteration of the natural evolutionary sequence for clusters with 
$W_{0}\geq 6.9$.

From the comparison with extragalactic GCs systems we have deduced that, 
in the case of LMC, SMC and Fornax system, these clusters have a Gaussian like distribution around a peak value $W_0\sim 5$. MW and M31 system are very similar in their features and $W_{0}$ distribution, with low-$W_{0}$ tail and a peak value in corrispondence of the stability limit. We can assume this as the main product of disk shocking, as well as extragalactic capture and low speed evolution objects mentioned above. For NGC5128 there is no low $W_{0}$ region tail, but only a narrow peaked distribution around the stability limit.

\end{multicols}

\begin{thebibliography}{99}
\bibitem{AH} Aguilar, L., Hut, P., Ostriker, J.P.: 1988, ApJ, 335, 720 
\bibitem{Ba} Barmby, P., McLaughlin, D.E., Harris, W.E., Gretchen, L.H., Forbes, D.A.: 2007, AJ, 133, 2764
\bibitem{BBBO} Bica, E., Bonatto, C., Barbuy, B., Ortolani, S.: 2006, AAP, 450, 105 
\bibitem{CH} Cohn, H., Hut, P.: 1984, ApJ, 277, L45 
\bibitem{Din} Dinescu, D.I., Terrence, M.G., van Altena, W.F.: 1999, AJ, 117, 1792
\bibitem{Dot} Dotter, A. et al.: 2010, AJ, 708, 698
\bibitem{GnO} Gnedin, O.Y., Ostriker, J.P.: 1997, AJ, 474, 223 
\bibitem{GGHRHW} G\`omez, M., Geisler, D., Harris, W.E., Richtler, T., Harris, G.L.H., Woodley, K.A.: 2006, AAP, 447, 877 
\bibitem{Hr} Harris, W.E.: 1996, AJ, 112, 1487
\bibitem{HuD} Hut, P., Djorgovski, S.: 1992, Nat, 359, 806
\bibitem{Kz} Katz, J.: 1980, MNRAS, 190, 497
\bibitem{K2} King, I.: 1966, AJ, 71, 64 
\bibitem{MG} Mackey, A.D., Gilmore, G.F.: 2003a, MNRAS, 338, 85
\bibitem{MG2} Mackey, A.D., Gilmore, G.F.: 2003b, MNRAS, 338, 120
\bibitem{MG3} Mackey, A.D., Gilmore, G.F.: 2003c, MNRAS, 340, 175
\bibitem{MG4} Mackey, A.D., Gilmore, G.F.: 2004, MNRAS, 355, 504
\bibitem{Qu} Quinlan, G.D.: 1996, NewA, 1, 255
\bibitem{VdB} Van den Bergh, S.: 2011, PASP, 123, 1044 
\bibitem{Zi1}  Zinn, R.: 1985, ApJ, 293, 424 
\end{thebibliography}
\end{document}